# Monitoring of Heart Movements Using an FMCW Radar and Correlation With an ECG

Rémi Grisot, Pierre Laurent, Claire Migliaccio, *Member, IEEE*, Jean-Yves Dauvignac, *Member, IEEE*, Mélanie Brulc, Camille Chiquet, and Jean-Paul Caruana

*Abstract*—Monitoring the activity of the heart is important for diagnosing and preventing cardiovascular diseases. The electrocardiogram (ECG) is the gold standard for diagnosing such diseases. It monitors the heart's electrical activity, and while this is highly correlated with the cardiac mechanical activity, it does not provide all the information. Other sensors such as the echocardiograph are able to monitor the heart's movements, but such tools are expensive and hard to operate. Therefore, contactless monitoring of the heart using RF sensing has gained interest in recent years. In this paper, we describe a process to extract the movements of the heart from millimeter wave radar with high accuracy, and thus we provide a noninvasive and affordable way to monitor cardiac movements. We then demonstrate the correlation between the observed movements and the ECG. Furthermore, we propose an algorithm to synchronize the ECG signal and the processed signal from the radar sensor. The results we obtained provide insights on the mechanical activity of the heart, which could assist cardiologists in their diagnoses.

*Index Terms*—Heart monitoring, FMCW, radar.

## I. INTRODUCTION

ACCORDING to the World Health Organization (WHO), "Cardiovascular diseases (CVDs) are the leading cause of death globally, taking an estimated 17.9 million lives each year" [1]. These diseases are a major public health issue in low- to middle-income countries [2] as well as in high-income countries [3] although the tools for prevention and diagnosis are more widely available in the latter. Heart disease is the leading cause of death in the United States, according to the Center for Disease Control (CDC) [3]. Prevention and early detection are critical if we want to reduce this mortality. Electrocardiography and echocardiography are the gold standards to monitor the activity of the heart. Electrocardiograms (ECGs) monitor the electric activity of the heart, whereas echocardiograms provide structural information about the heart and blood vessels. These diagnostic tools enable detection of cardiovascular diseases. However, they are expensive and require training to operate. Frequency Modulated Continuous Wave (FMCW) radar is a type of sensor that is very sensitive to small movements. It could provide a complementary method to ECGs and echocardiography, for some pathologies, that is cheaper and easier to operate. Although, at the frequencies used (77-81GHz), the waves emitted by the radar only penetrate the skin a couple millimeters [4], [5], we are able to detect skin movements induced by the heart beats. As the interpretation of ECGs is widely documented and the P, Q, R, S, and T waves are linked to heart movements, correlating the movements detected by the FMCW radar sensor with the ECG signal would allow us to benefit from the knowledge of cardiology from the ECG in interpreting the radar sensor signal. The monitoring of vital signs using FMCW radar sensors has gained tremendous interest over the past several years [6], [7], [8], [9], [10], [11], [12], [13], [14], [15], [16], [17]. However the monitoring proposed is often focused on detecting the heart rate over a period of time whereas we aim to detect each heartbeat individually and explain each part of the movement. Some previous work has shown it is possible to reconstruct the ECG from the radar signal, using neural networks [18], [19]. One drawback of such an approach is that the neural network is trained to produce a normal ECG, as it is trained on data from healthy subjects. Hence, the results it may produce for subjects with CVDs is unpredictable. Moreover, some pathologies like heart failure have an impact on the mechanical activity of the heart, but are barely detected using an ECG. Therefore, in this paper, we do not try to reconstruct the ECG from the radar signal. Instead we demonstrate the correlation between the two signals and explain it from a medical point of view. The use of the ECG helps us explain the movements we observe because the link between the electrical and the mechanical activities of the heart is well known [20]. Furthermore, because we do not use a neural network to compute our signals, as in [21], our whole process is less computationally expensive and does not require training. Pioneering work was carried out in [22], where the authors showed some variations in the movement correlated to the electrical activity recorded by an ECG, however the observed correlation can be improved. While they used a Doppler Radar System (DRS) at 24 GHz, we chose to use a FMCW radar at a higher range of frequencies. Hence, we have a greater resolution in velocity. In our signal, the peaks in the velocity signal have a prominence that is greater by three orders of magnitude, which makes them easier to detect and Rémi Grisot is with the Laboratoire d'Electronique, Antennes et Télécommunications (LEAT), Université Côte d'Azur, 06903 Sophia Antipolis, France, and also with XtreamWave, 83000 Toulon, France (e-mail: remi.grisot@xtreamwave.com).

Pierre Laurent is with the Hôpital Sainte Musse, 83100 Toulon, France.

Claire Migliaccio and Jean-Yves Dauvignac are with the Laboratoire d'Electronique, Antennes et Télécommunications (LEAT), Université Côte d'Azur, 06903 Sophia Antipolis, France.

Mélanie Brulc is with XtreamWave, 83000 Toulon, France, and also with UMR CNRS, IM2NP-ISEN, 83000 Toulon, France.

Camille Chiquet and Jean-Paul Caruana are with XtreamWave, 83000 Toulon, France.

interpret. We also have a better resolution in range due to the wider bandwidth.

In our paper, we propose an experimental setup to collect data simultaneously from an ECG and an FMCW radar. Then we describe our clustering-based method to detect P, Q, R, S, and T waves on an ECG signal and our process to extract cardiac-induced motion from the radar signal. This process offers a good tolerance to missing data and allows the torso movements induced by the heart to be monitored with a high accuracy. Finally, we establish a correlation between the data from the two sources, we demonstrate it with an experiment and we propose an algorithm to automate this synchronization. The signals extracted and interpreted in this paper have been shown to be reproducible among the subjects that took part in the experiment.

## II. Mechanics of the Heart

As the radar detect the movements of the chest induced by the heart, it is important to understand the different steps of the cardiac cycle from a mechanical perspective. We explain below the basics of heart movements and their link to the ECG as described in [20]. Fig. 1 shows the main cavities of the heart and the parts involved in the conduction of the depolarization through the heart.

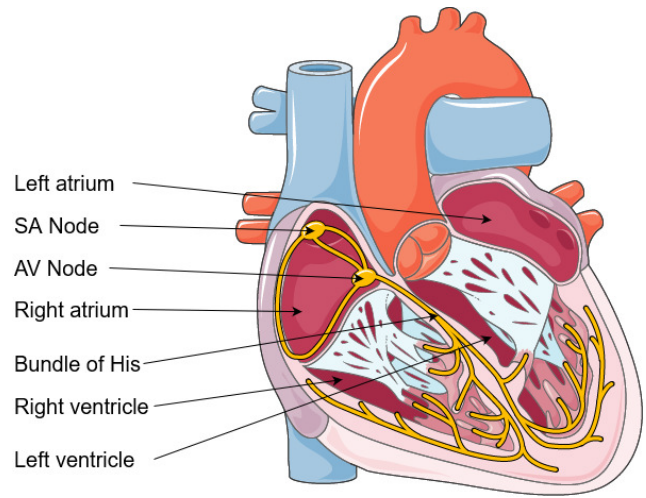

Fig. 1.  Schema of the myocardium.

The myocardium, i.e., the cardiac muscle, can be separated into two blocks: the atria (left and right) and the ventricles (left and right). For a healthy subject, a positive electrical impulse (depolarization wave) is generated in the sinoatrial node (SA node), in the right atrium. As the depolarization wave propagates through the atria (P wave on the ECG), they contract and expel the blood they contain into the ventricles: this is the atrial systole. The depolarization wave reaches the atrioventricular node (AV node) and is slowly conducted through it. But when it reaches the ventricular conduction system, it spreads quickly through the right and left branches of the bundle of His. This leads to the depolarization of the right and left ventricles (QRS complex on the ECG) and to their contraction. Thus, the QRS complex corresponds to the initiation of the ventricular systole (i.e., contraction). The repolarization of the ventricular myocytes lasts from the end of the QRS complex to the end of the T wave and occurs in two steps. The ventricular systole ends with the T wave. As the ventricles are bigger cavities (hence more muscle surrounds them) than the atria and their depolarization occurs faster than the depolarization of the atria, the ventricular systole induces a movement of the myocardium that is faster and has a greater amplitude than the movement induced by the atrial systole. The different steps of the cardiac cycle and their correspondence with ECG waves are represented in Fig. 2.

FMCW radar sensors allow us to monitor small displacements. With the frequencies that we use (77-81 GHz), the waves emitted by the radar do not penetrate the skin more than a few millimeters [4], [5]. Thus, they do not reach the heart itself. However, the different phases of the heart cycle are composed of rapid, strong contractions and releases that induce small movements of the skin of the monitored subject. As explained in [19], there is a transfer function that links the heart movements to the observed skin movements. While this transfer function is not known, we can infer the movements that we expect to observe. Indeed, the biggest and fastest of the movements that comprise the cardiac cycle is the ventricular systole. During this step, the heart goes from its maximum overall volume to its minimum one in a short time. Thus, we expect the skin to quickly move away from the radar during the ventricular systole.

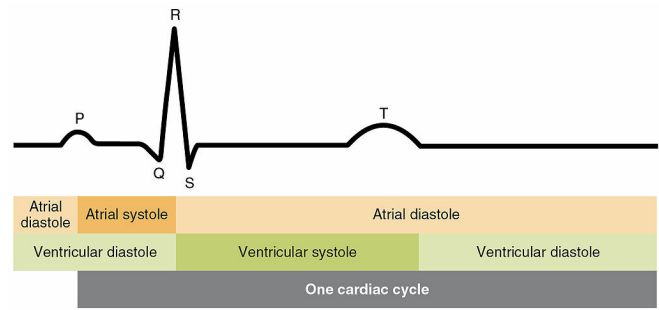

Fig. 2.  ECG waves and association with cardiac cycle [23].

## III. Data Collection

### A. Overall Methodology

We follow the steps described in Fig. 3. As the signals are of different natures, because they are collected from different sensors, we use different processing techniques on them. When the data flows re-join, they are not synchronized. As described below, there is an operator-induced offset in one of the signals. When must then synchronize the data to be able to interpret it. To achieve this we use an approach based on cardiology knowledge, and more specifically on the mechanics of the heart, and we setup an experiment to verify it.

### B. Experimental Setup

The purpose of this experiment was to demonstrate the correlation between the ECG and the radar signal. Therefore, the setup described below corresponds to the conditions that were determined to be optimal for observing the heart-induced motions. The experiment consisted of the simultaneous recording of an ECG and an FMCW radar (captures) for two setups and for three configurations. For each configuration, two captures were made for each subject. The recordings were launched by an operator on two different computers. Therefore, the signals might not be perfectly synchronous. We will describe in a later section our synchronization protocol. The experiment was performed on 22 subjects: 11 males and 11 females. The subjects were 20 to 58 years old, with an average age of 25.5. They all provided informed consent to take part in this experiment. In the first part of the experiment, the subject was in a supine position as can be seen in Fig. 4 left (setup #1). The radar sensor was positioned horizontally above the chest, in a fixed position, at a distance of about 50 cm. Subjects were asked not to move or talk, and to breathe slowly. We then collected data from the two sensors for one minute (first configuration). Then we asked them to hold their breath with their lungs full and we collected data for 30 s (second configuration). And finally subjects were told to hold their breath with their lungs empty for another 30 s data collection (third configuration). Then, subjects lay in a right lateral decubitus position as can be seen in Fig. 4 right (setup #2) and we repeated the 3 capture sessions (slow breathing, apnea with full lungs, apnea with empty lungs). The protocol uses these two setups (supine and right decubitus) for two reasons: the first one is that the cardiac movement is complex and multi-directional. Hence, the displacement induced by the heartbeat in the different directions is not equal. Thus, having a different angle of view provided additional insight about the movement. The second reason is that the aorta, between the heart and the lower abdomen, lies very close to the skin and is in direct sight of the radar in the supine position. It might then produce a bigger movement than the heart in terms of skin displacement. Correlating and comparing the movements seen from the two positions gives a good overview of the heart kinetics. The two setups used for the experiment (subjects lying on their back or their side) are represented in Fig. 4. The apnea captures were added to the protocol because we noticed that the breath had an impact, not only by modulating the frequency of the heartbeat due to physiological sinus variation activity [24], but also on our ability to detect the beat properly. Indeed, as we are detecting the skin movements induced by the heartbeats, it appears that the skin displacement is not the same depending on whether the chest is inflated or not.

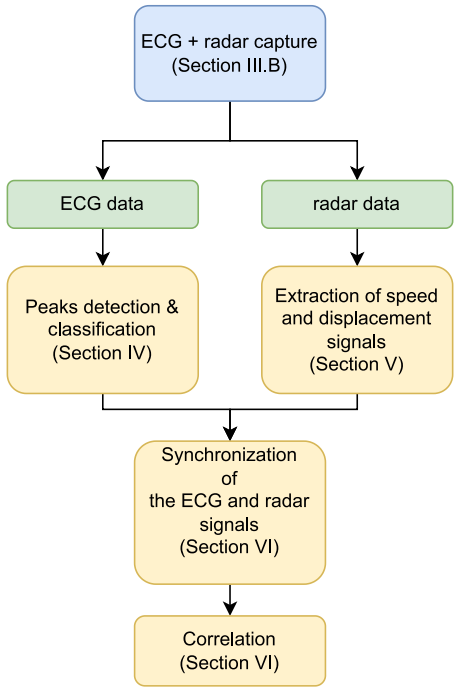

Fig. 3. Block diagram of the steps we followed to process and correlate the signals from the ECG and from the radar.

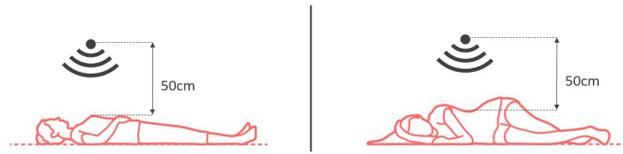

Fig. 4. The left image shows the subject and radar positions in the first part of the experiment (supine position); the right image shows the positions in the second part of the experiment (right lateral decubitus).

### C. Configuration Details

*1) ECG:* For the experiment we used a 6-Lead ECG, with the 4 electrodes placed on the wrists and ankles. We collected one electrical signal for each lead. The collected data from the ECG had a sampling frequency of 2 kHz. We had three bipolar derivations: I, II, III and three augmented derivations: AVR, AVL, AVF. Each derivation can be seen as a way to look at the heart, with a certain angle. The signal obtained for each bipolar derivation corresponds to the difference in potential between two electrodes. The Lead I signal is the difference between the potentials of the right arm and left arm; Lead II is the difference between right arm and left foot and Lead III is the difference between the left arm and left foot. The augmented derivations are computed from the bipolar derivations:

$$AVL = \frac{Lead_I - Lead_{III}}{2}$$
$$-AVR = \frac{Lead_I + Lead_{II}}{2}$$
$$AVF = \frac{Lead_{II} + Lead_{III}}{2} \quad (1)$$

The obtained derivations and their corresponding angle of view can be seen in Fig. 5.

*2) Radar:* The radar we used is a IWR1642Boost [26] from Texas Instruments®, working in the frequency range [77-81]GHz. This radar is a MIMO radar that we use as a SIMO radar with 4 Rx antennas. The antennas are spaced by $\frac{\lambda}{2}$ (computed at the central frequency), which provides a field of view of 180° without grating lobes. For all records we used the whole 4 GHz bandwidth. Hence, we have the smallest possible range resolution for that radar, which is $d_{res} = \frac{c}{2B} = 3.75 \times 10^{-2}m$, with $c$ the speed of light in meters per second, and $B$ the bandwidth in Hertz. The phase noise of the IWR1642Boost is $-93\ dBc/Hz$, which is compatible with the range resolution we want to achieve.

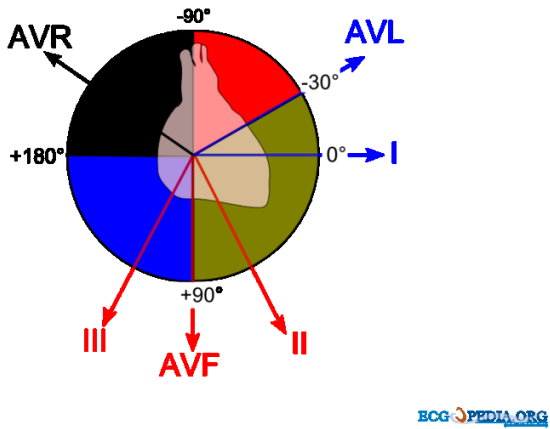

Fig. 5. ECG leads and associated angles. We had three bipolar derivations and three augmented ones [25].

The distance of 50 cm between the radar and the torso of the subject was chosen as a compromise. On the one hand, getting closer to the sensor increases the phase difference between the signal received on the different antennas. This has an impact on the summation operation described in section V-C, adding some noise. The phase difference between the signals received by two Rx antennas spaced by $\frac{\lambda}{2}$ can be seen in Fig. 6. On the other hand, if we increase the distance between the radar and the subject, we become more sensitive to perturbations induced by the environment. As the radar field of view is 180°, anything that is at the same range as the target will interfere with the heart signal we want to extract.

At the output of the radar sensor, we collect the IQ data directly, as an output of the Analog to Digital Converter (ADC). To get access to that raw information, we use the radar paired with a DCA1000 board. The DCA1000 board receives raw IQ data from the radar via an LVDS interface and transmits them to the operator's computer via an Ethernet interface. UDP is used to communicate on the Ethernet interface, as is often the case for time sensitive applications. However, unlike with TCP, there is no guarantee that the data will be correctly transmitted between the emitter and the receiver. Thus, some data packets may end up missing on the operator's computer. In our case, missing data were replaced with zeros. With the IWR1642Boost radar device, groups of subsequent chirps are sent. Those groups are called frames. All the frames are equally spaced in time, and within each frame, the chirps are regularly spaced too. But the inter-frame time is greater than the inter-chirp time, thus creating a time discontinuity between the last chirp of the $(n-1)^{th}$ frame and the first chirp of the $n^{th}$ frame that we had to deal with. This inter-frame time is caused by the computing performed by the device on each frame, which cannot be disabled. This time discontinuity can be observed in Fig. 7. In our setup, we have 625 frames per second and each frame contains 16 chirps. For each chirp, we have 64 samples, i.e., 64 IQ values. As we want the best possible resolution in velocity, we use the shortest chirp time allowed by the radar. This corresponds to a slope of 100 MHz/µs.

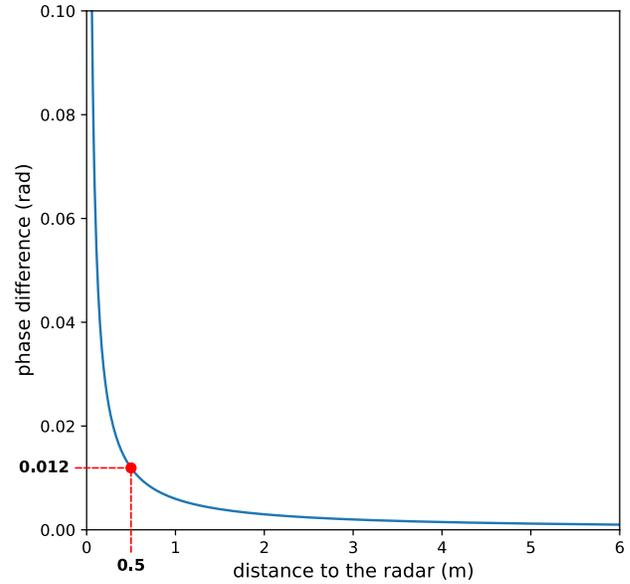

Fig. 6. Phase difference of the signal received by two antennas spaced by half the wavelength, as a function of the distance of the target. The distance chosen for the experiment was 0.5 m.

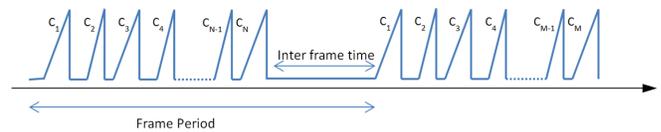

Fig. 7. Discontinuity of time between chirps of successive frames. Source: Texas Instruments®.

## IV. ECG Data Processing

### A. Filtering

As our goal was to correlate the radar signal with the ECG waves, we processed the ECG data to extract the positions of the P and T waves and the QRS complex. Therefore, we kept only the signal from derivation II as it provides a good view of these waves. We applied three layers of filters to the ECG signal, as explained in [27]: a high-pass filter with a cutoff frequency of 0.05 Hz, to remove very low frequency (VLF) components that are at sub-respiratory frequencies; a low-pass filter with a cutoff frequency of 75 Hz to remove very high frequency (VHF) noise; a band-stop filter, with a stop band from 45 Hz to 55 Hz, to remove noise caused by power-line interference. After having applied these three layers of filtering, we still observed baseline wander, which makes the detection of the ECG waves difficult. To extract the baseline, we computed the second order moving average on the filtered signal, with a window of 2001 samples, which corresponds to 1 s of data. Finally we subtract the extracted baseline from the filtered signal. Hence we obtain a filtered signal with a constant baseline (see Fig. 8 top).

### B. P, Q, R, S, and T Wave Detection

In the processed ECG signal, we wanted to detect the P, Q, R, S and T waves. The Q and S waves are the bases of the R peak. Thus, we focused on detecting only the P,

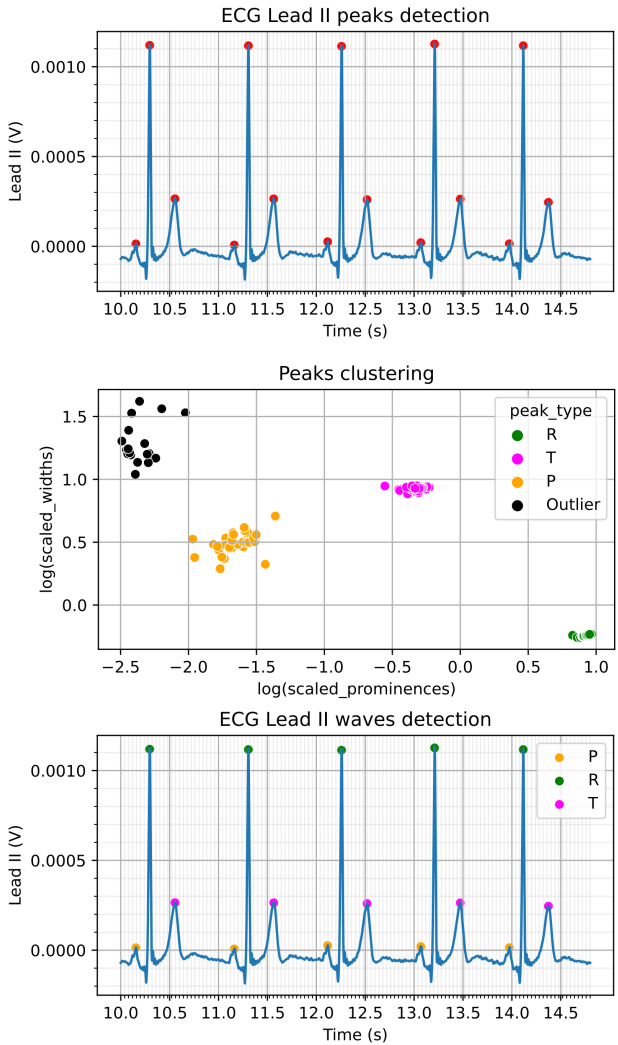

Fig. 8. The top graph shows the result of peak detection applied to the Lead II track. The middle graph shows the result of the DBScan clustering applied to the previously detected peaks based on their prominence and width. The bottom graph shows the labeling provided by the clustering algorithm applied to the originally detected peaks.

R and T waves. The wave detection occurred in two parts: the detection of all peaks in the signal and the subsequent classification of those peaks. First, for the detection part, we used a *find_peaks* algorithm, provided by the *scipy* [28] package in Python. We considered a peak if its prominence was greater than a fifth of the standard deviation of the signal and if it lasted less than 200 ms. These parameters were found to detect all peaks of interest while minimizing the number of false detections. Then, we needed to classify these peaks into 3 classes: P-wave, R-wave and T-wave. One of the main challenges of classifying these waves is that their properties, such as peak prominence or width, can vary greatly from one individual to another. Hence, finding general properties to classify detected peaks is difficult. To address this problem, we used a clustering-based approach. For each ECG capture, we applied a clustering algorithm to the data: DBScan. The features used for the clustering were the peak prominence and the peak width. We chose DBScan as the clustering algorithm because of its density-based approach and because it can detect outliers. Thereby, we could discard any possible false detections by the *find_peaks* algorithm. As the features in our dataset were on different scales (hundreds of ms for peak duration; mV for peak prominence), we scaled our data by the standard deviation. Furthermore, in order to better separate the clusters, especially for values near zero, we took the logarithm of the scaled data. The result of the clustering is shown in Fig. 8 (middle). Once the clusters were separated, we labeled them using the wave properties. The R waves are the most prominent and are short in time, while the T waves have the longest durations. Then, we added a last verification step, where we ensured that the wave order was respected (P followed by R, followed by T, Fig. 8 (bottom)).

## V. RADAR DATA PROCESSING

### A. Data Format

As noted above, we recorded the IQ signal just after the ADC. We reorganized it in a 4-dimensional matrix of complex numbers: [Rx antennas, frames, chirps, samples]. In order to have equally spaced points in the time dimension (carried by both the frames and chirps dimensions of our matrix), we kept only the first chirp of each frame, in order to avoid the apparent gap between chirps introduced by the inter-frame time. Thus we had a 3-dimensional matrix to process. The "samples" dimension is the fast time dimension, whereas the "frames" dimension is the slow time dimension.

### B. Range Selection

The first step consisted of applying a Fast Fourier Transform (FFT) along the samples dimension. The modulus of the result gives the intensity of the reflected signal for the corresponding frequency bins (as the FFT is discrete), which can be easily converted to distance bins using the formula $d = f \times \frac{c \times r}{2 \times B}$ where $f$ is the intermediate frequency associated with the frequency bin in Hz, $c$ is the speed of light in $m.s^{-1}$, $r$ is the ramp time in seconds and $B$ is the bandwidth in Hz. We can see the evolution of the modulus of the range FFT in Fig. 9. Using this result, we could focus on the range index corresponding to the highest intensity, which we will refer subsequently as the brightest range. To find this range, we computed the mean modulus of the range FFT over time for each range index. We obtained plots like that shown in Fig. 10. Using $argmax$, we then selected the range index with the highest mean modulus. This range index corresponded to a distance around 50 cm of the radar, which agrees well with the distance between the radar and the observed subject at azimuth 0°.

### C. Velocity Extraction

$$STFT(s)[f, m] = \sum_{k=0}^{n-1} s[k+m] \times W[k] \times e^{-j.2\pi.k.\frac{f}{n}} \quad (2)$$

where $n$ is the number of frequency bins of the FFT and $m$ is the frame index. The squared modulus of the result gives us a spectrogram, showing the frequency shift of the target with time. By carefully choosing the window size and the number of bins, we could reach a high resolution in both time and frequency. We used a Hann window as windowing function to soften the artifacts at the edges of the window during the FFT, and a sliding step size of 1 in order to preserve the temporal resolution. Using such a window also widens the main lobe, which is not a concern here, as we are only interested in the index of the tallest peak, not its properties. The resulting spectrograms are like the one shown in Fig. 11. We converted the frequency bins obtained into velocity bins (in $m.s^{-1}$) by applying the formula: $v = f \times \frac{\lambda}{2}$, where $v$ is the radial velocity in $m.s^{-1}$, $f$ is the intermediate frequency associated with the frequency bin in $Hz$ and $\lambda$ is the wavelength corresponding to the central frequency ($\lambda = 3.8\,mm$ in our case). We reached a temporal resolution of $1.6\,ms$ and a velocity resolution of $1.45 \times 10^{-4}\,m.s^{-1}$.

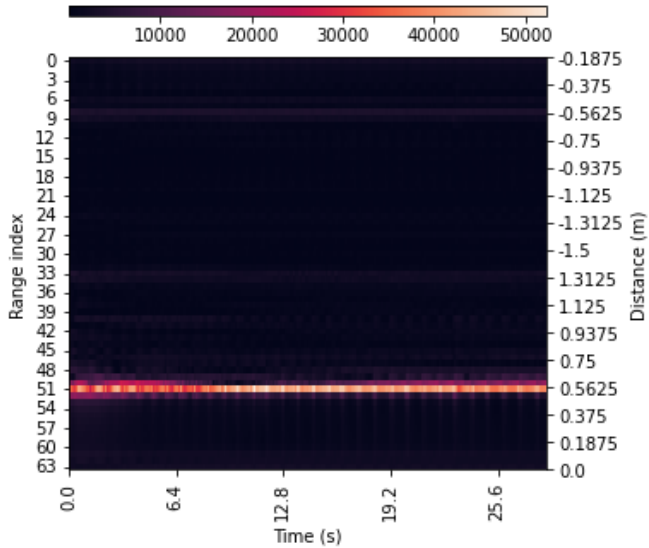

Fig. 9. Example of a heat-map of the range FFT modulus obtained for one of the Rx antennas (sorted by frequency range index). The frequency range index corresponds to the index of the frequency bins obtained from the discretization that occurs during the FFT computation.

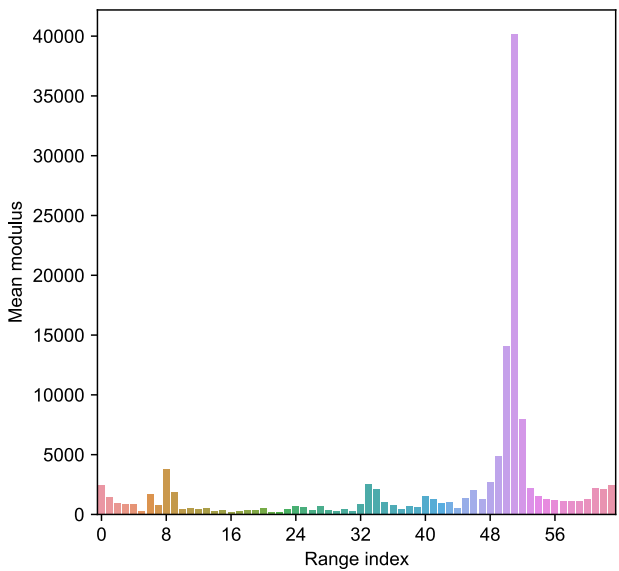

Fig. 10. Mean intensity for each range computed for the data shown in Fig. 9. We can see that the range at index 51 has a much higher intensity on average than the others.

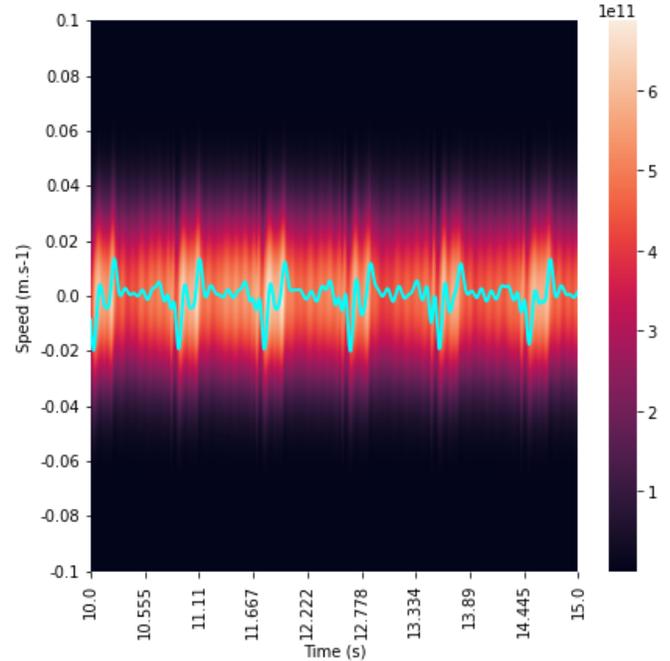

Fig. 11. Spectrogram obtained for one antenna, at the brightest range (zoom on 5 s). The extracted heart signal (scaled by a factor 30 to be visible) is superimposed (cyan curve).

We then summed the signal of the spectrograms obtained for the four Rx antennas, for each range. Finally, we took the *argmax* along the velocity dimension for each time step and for each range. By converting the frequency bin into a velocity we determined, for each range, a time-series of velocity. We passed our signals through two layers of moving average, one with a small window to remove high-frequency noise and one with a large window to extract macro-movements from the signal to subtract them from the original signal. We obtained curves of velocity versus time, an example is shown in Fig 12. Regular down peaks are observable and if we superimpose and manually synchronize the vertical lines corresponding to the R-peaks on the ECG to match with these downward peaks, we can see that the period matches perfectly between the two signals. In Fig. 13, we show a zoom on 5 s of the velocity signal, with the ECG superimposed in red and the vertical green and magenta lines corresponding to R and T waves, respectively. We can see there is a recurring pattern in the radar signal and that not only do the downward peaks match each R peak on the ECG but also a smaller upward peak matches the T wave peak. The hypothesis used for the manual synchronization between the two signals is explained in the following section.

We observed that there are no significant differences between the captures realized in apnea with inflated lungs and those realized in apnea with empty lungs. However, the subjects found more difficult to hold their breath when their lungs were empty. Sometimes, as they were struggling to hold their breath, they made some uncontrolled movements. Hence, we recommend to realize apnea capture with full lungs only, in order to have data of better quality.

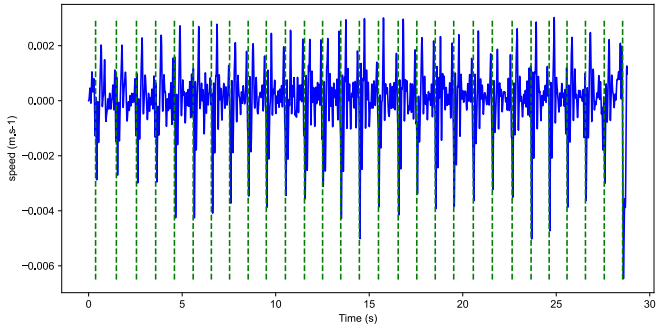

Fig. 12. Evolution of velocity with time, for the brightest range (blue curve). The superimposed vertical dashed green lines correspond to the times where the R-peaks are detected on the ECG (manual synchronization).

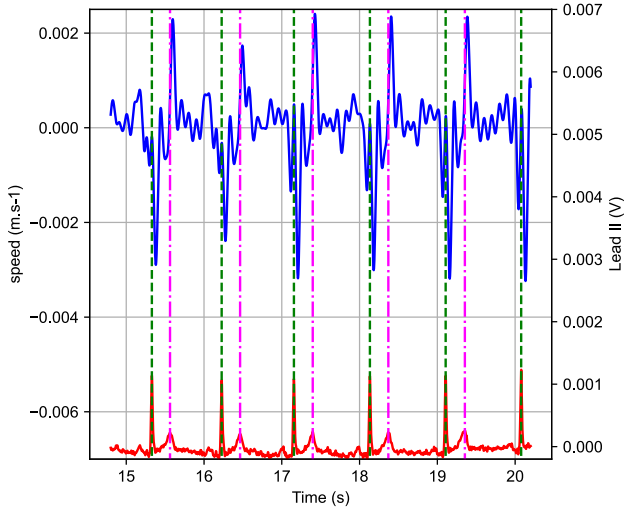

Fig. 13. Evolution of velocity with time (blue curve), for the brightest range (zoom on 5 s). The superimposed vertical dashed green lines correspond to the times where the R-peaks are detected on the ECG (manual synchronization). The superimposed vertical dashed magenta lines correspond to the times where the T-peaks are detected on the ECG. The red curve represents the processed signal from the Lead II track of the ECG.

TABLE I
COMPARISON OF OUR PAPER WITH OTHER PAPERS
MONITORING VITAL SIGNS

| Ref. | Vital sign | Radar / Method | ECG Comparison |
|---|---|---|---|
| [11] | Breath | FMCW / Spectrogram, envelope detection | NA |
| [19] | Heartbeat | FMCW / Phase extraction, Neural Network | + |
| [22] | Heartbeat | DRS / RSC | + |
| This paper | Heartbeat | FMCW / velocity extraction | ++ |

### D. Discussion and Comparison of Results

Table I summarizes recent examples of vital signs monitoring using radar and how our work compares with them. In this section, we compare our results with those obtained in [11], [19], and [22]. First, in [11], they use envelope detection to extract the velocity of the subject's torso, in order to monitor breathing. While this method provides accurate detection of the breath, it is not precise enough to recover the movements of the heart. A major difference between our method and the ones presented in [19] and [22] is that our method allows us to extract the velocity directly while the methods in [19] and [22] extract the position. We can easily switch from one signal to the other as the speed is the derivative of the position. However, if the input signal contains errors or noise, it is safer to integrate than to compute the derivative. Abrupt variations in the signal will result in high peaks in the derivative. In contrast, integrating a signal is an operation that is less sensitive to noise. Concerning [19], we compare our results to the data they obtain before they feed it into their neural network. This input corresponds to a displacement time-series (i.e., a mechanical signal) whereas the output of their neural network is the reconstruction of an ECG, i.e., an electrical signal. Thus it makes more sense to compare the two mechanical signals. In [19], the displacement signal is obtained by extracting the phase of the range FFT over time, at a given index. Computing the derivative of this signal gives us a velocity time-series. This method is subsequently referred to as the "phase method". First, we can see from Fig. 14 that if we compute a point-to-point distance between the curves obtained with the "phase method" and with our "argmax method", there is an average difference of $1.8 \times 10^{-4} m.s^{-1}$. This small average distance indicates that the two curves are very much alike, so there is no big difference in terms of resolution in velocity. The "phase method" has the advantage that it requires far less calculation than the argmax method. As explained above, the "argmax method" requires the calculation of an STFT which is quite computationally expensive if we want to achieve a good resolution.

We now compare the two methods regarding their robustness to data loss. As we explained in section III-C2, some packets may be lost between the radar sensor and the operator's computer, mostly due to the use of UDP. In order to measure the robustness of the two methods regarding missing data, we randomly removed (replaced with zeros) 1% of the data in a clean signal (without loss). We chose the value of 1% because it is the worst case we observed during our measurements. We then extracted the velocity time-series using the "phase method" and the "argmax method". The results can be seen in Fig. 15. We observe that the "phase method" is much more sensitive to missing data. Missing data induce high peaks in the velocity time-series as high as $0.2\ m.s^{-1}$, which is 40 times the height of the big downward peaks corresponding to heartbeats that we observe in the clean signal. The average signal-to-noise ratio (SNR) for the "phase method" is $-23.2972\ dB$ whereas it is $11.2541\ dB$ with the "argmax method". This shows that our method is much more resilient to data loss.

In [22], the authors used the Random Sample Consensus (RSC) algorithm to compute the position from their data. Even if this algorithm is robust, it is non-deterministic. Hence, there is only a given probability that the obtained result will be correct. Even if this probability can be high, it is never 100%. This uncertainty may induce additional noise in the data. Our method is deterministic, making it more robust. We measured the evolution of the RR interval for the ECG and the evolution of the interval between two big down peaks for the radar (Fig. 16). Such an analysis was done in [22]. Hence, we can compare our work to theirs based on the metric they have chosen. There is a documented variability in the duration of the cardiac cycle. Such variability is visible on the ECG (by monitoring the RR interval). We observe the same variability with our method (monitoring of the interval between two successive downward peaks). We can observe that with our method, the difference between the interval measured between two R peaks on the ECG and the interval between two down peaks on the velocity signal extracted from the radar is much smaller. Considering the R-R interval as the target and the interval between peaks on the velocity signal as the prediction, they obtain a Root Mean Squared Error (RMSE) of 14.67 ms, on one subject. With our method, we get a RMSE of 0.0217 ms, computed on 12 subjects. This shows again that our method is more robust and less prone to error in the detection of the ventricular systole, as the corresponding peaks in the radar velocity signal are tall and easily detectable.

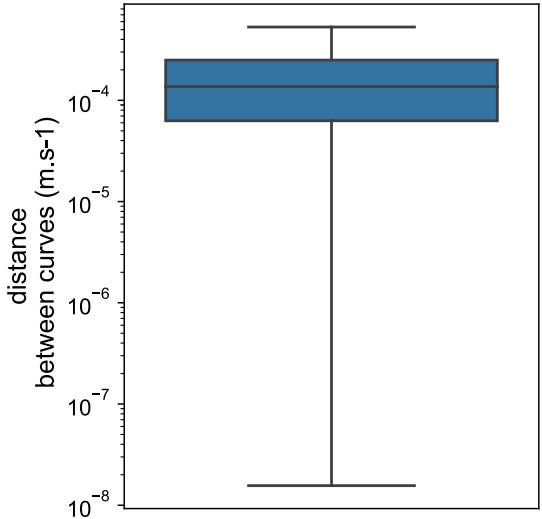

Fig. 14. Point-to-point distance between the velocity time-series extracted with the phase method and the velocity time-series extracted with our method. Logarithmic scale on the y axis.

## VI. SYNCHRONIZATION OF ECG AND RADAR SIGNALS

### A. Based on Cardiology Knowledge

As stated previously, the two captures are launched separately, on two different computers, by an operator. Such a setup can induce an offset between the two signals and we have to manually synchronize them. Because the velocity alone might be difficult to apprehend (for example, a succession of small peaks with the same orientation can lead to the same displacement as a single big peak), we converted the velocity into a displacement by integrating it. The results are shown in Figures 17 and 18. The link between the electrical activity and the mechanical activity of the heart is known [20], as exemplified in Fig. 2 and described in section II. The QRS complex corresponds to the ventricular depolarization and initiates the ventricular systole. The ventricular contraction is the fastest and biggest movement of the cardiac cycle. Hence, we expect that movement to correspond to a drop in the displacement signal and a big downward peak in the velocity signal, because when the ventricles contract, the heart moves away from the radar, inducing a negative velocity. We also know that the ventricular repolarization, which leads to the relaxation of the ventricles, occurs after the S wave in two phases and lasts until the end of the T wave. Furthermore, we know that the repolarization is slower than the depolarization. Consequently, we can associate the repolarization of the ventricle with the succession of upward peaks in the velocity signal and the two-step rise in the displacement signal. A similar synchronization hypothesis was made in [22]. The result of a manual synchronization based on this hypothesis is shown in Figures 13 and 18.

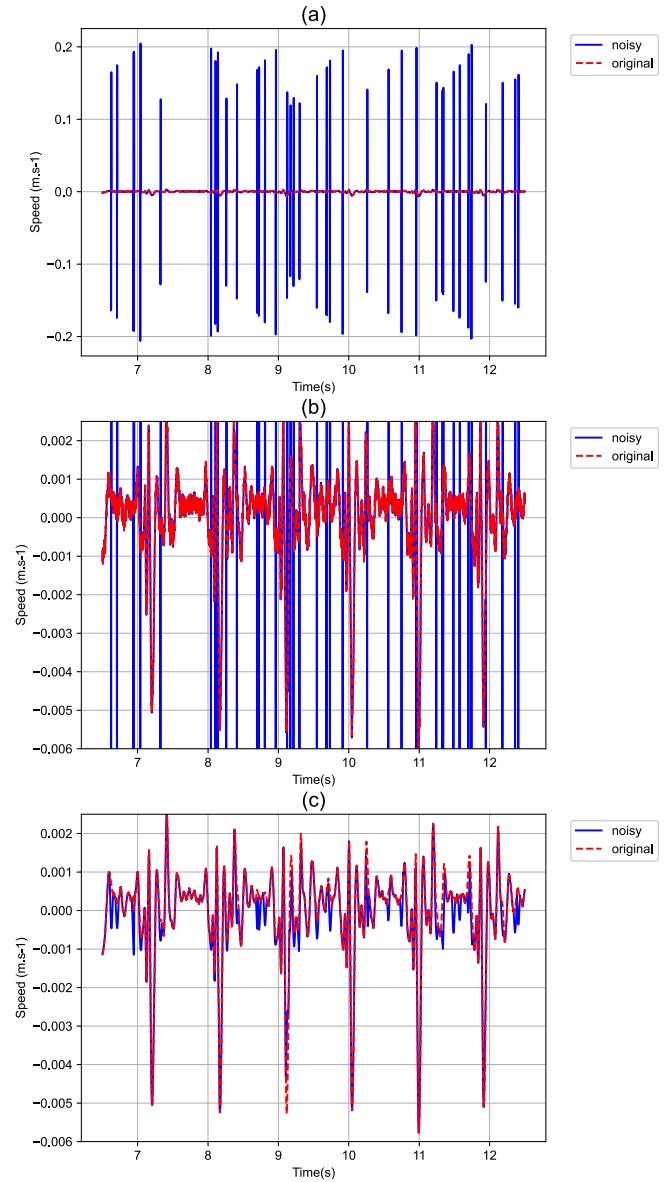

Fig. 15. Velocity extracted with various methods from data with 1% missing packets (blue curve) and comparison with the velocity extracted with the same method from the same signal but with no missing packets (red dashed curve). Graph (a) shows the velocity extracted from the phase of the range FFT, as in [19]. Graph (b) is a zoom on graph (a), to focus on the heart signal. Graph (c) shows the velocity extracted with our method.

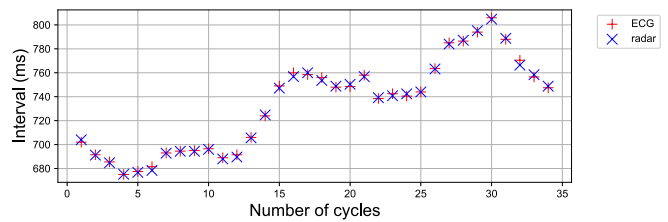

Fig. 16. Evolution of the R-R intervals on the ECG signal (red dots) and of the intervals between big downward peaks on the radar signal (blue dots) in a 30s capture.

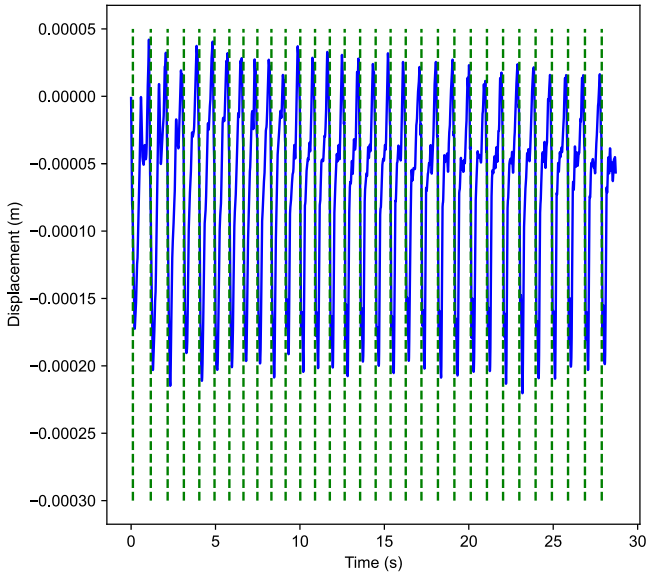

Fig. 17. Evolution of the chest displacement with time (blue curve), for the brightest range. The superimposed vertical dashed green lines correspond to the times where the R-peaks are detected on the ECG (manual synchronization).

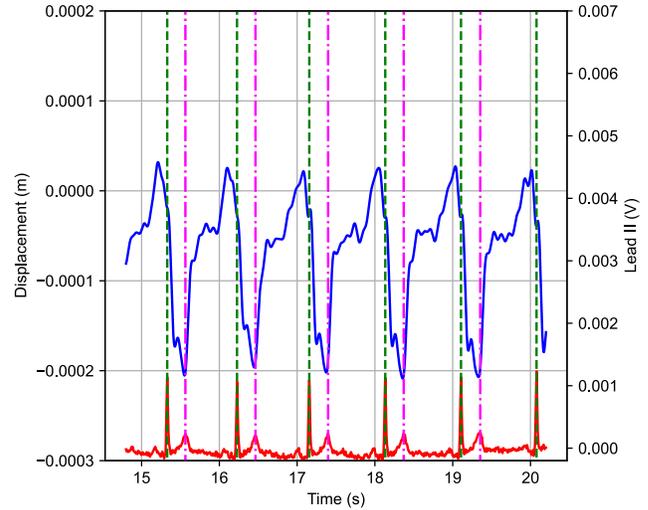

Fig. 18. Evolution of the chest displacement with time (blue curve), for the brightest range (zoom on 5 s). The superimposed vertical dashed green lines correspond to the times where the R-peaks are detected on the ECG (manual synchronization). The superimposed vertical dashed magenta lines correspond to the times where the T-peaks are detected on the ECG. The red curve represents the signal processed from the Lead II track of the ECG.

### B. Experimental Verification

In the previous section, we presented a hypothesis, based on knowledge of the cardiac cycle and its links with the ECG, for synchronizing ECG and radar signals. In this section, we describe an experiment to validate our hypothesis. As we previously stated, the capture on the two sensors was launched independently and manually by an operator. While he tried to do it synchronously, we cannot guarantee that there was no delay between one capture and another. In order to perfectly synchronize the two signals, we have to detect, on one of them, when the capture starts on the second sensor. It is easier to make an observable and controllable perturbation on an ECG signal than on a radar signal because it is an electrical signal.

*1) Experimental Setup:* For the experiment, we set up the circuit shown in Fig. 19. The computer controlling the radar sensor was connected to a microcontroller (e.g., an Arduino board). When we launched the capture on the radar, a signal was sent via USB to the microcontroller. When it receives the signal, the microcontroller set a given GPIO (General Purpose Input/Output) to the UP state for 100 ms then it set it to the DOWN state again. The GPIO was connected to the base of a transistor. When the GPIO was in the UP state, the potential $V_{dd}$ was applied to the transistor emitter. The emitter was connected to the right arm lead of the ECG. This meant that the perturbation would be visible on both tracks I and II. For the experiment, the potential $V_{dd}$ was 3.3 V. When we apply such a voltage, we expect to see a huge up front followed by a similarly huge down front, because a normal ECG signal typically has a magnitude of only a few mV. Such fronts can be easily detected. The up front was detected by using a threshold on the first derivative of the signal of one of the concerned leads (either Lead I or II). The threshold was chosen so that the impulse was detected in one capture for one subject, both randomly selected. The appropriateness of the chosen value was then confirmed as the impulse was successfully detected in all the other captures of the experiment. An example of such a detection can be observed in Fig. 20. We repeated the experimental setup described in section III (captures with slow breath or apnea in supine and right decubitus position), but with the electrode connected to the circuit of Fig. 19. In order to ensure that we would detect the perturbation in the ECG, we began recording the ECG 3 s before the radar. This experiment was performed on a smaller number of subjects (3 subjects, 2 males, 1 female): its purpose was to validate our synchronization hypothesis so that we could use it to synchronize the data recorded without the impulse on the ECG.

*2) Synchronization:* Once we detected the perturbation, because we know that it coincides with the time of the start of the radar capture, we could apply a shift to the ECG time to ensure the $t_0$ of both signals corresponded. The signal before the perturbation was discarded. Finally, in order to remove the perturbation from the ECG signal, we removed 1 s of data at the beginning of both signals. The resulting signals were then processed in the same way as described in the previous sections.

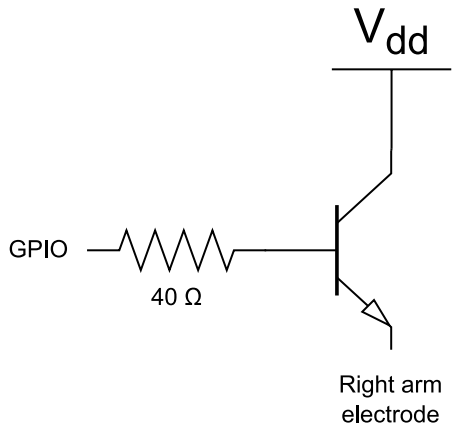

Fig. 19. Schema of the electrical circuit used to add a visible perturbation to the ECG signal when the radar capture starts.

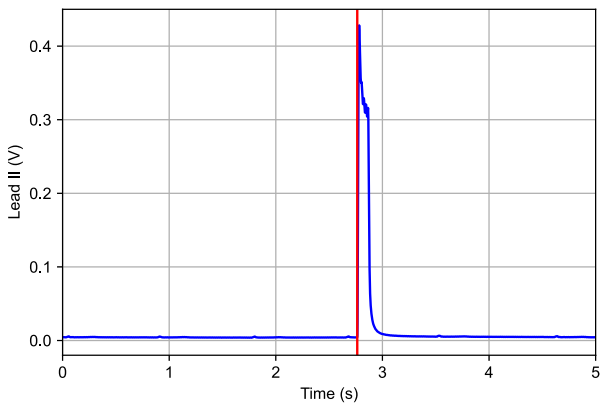

Fig. 20. Detection of the impulse sent on the right arm electrode when the radar capture started. The blue curve is the Lead II track of the ECG. The vertical red line shows the time when the impulse was detected.

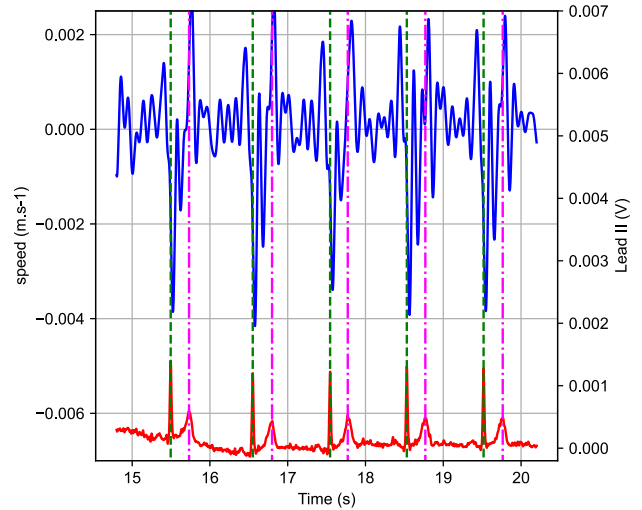

Fig. 21. Evolution of velocity with time, for the brightest range. The radar signal was synchronized with the ECG using the impulse experiment. The subject is the same as in Fig. 13. The superimposed vertical dashed green lines correspond to the times where the R-peaks are detected on the ECG (manual synchronization). The superimposed vertical dashed magenta lines correspond to the times where the T-peaks are detected on the ECG.

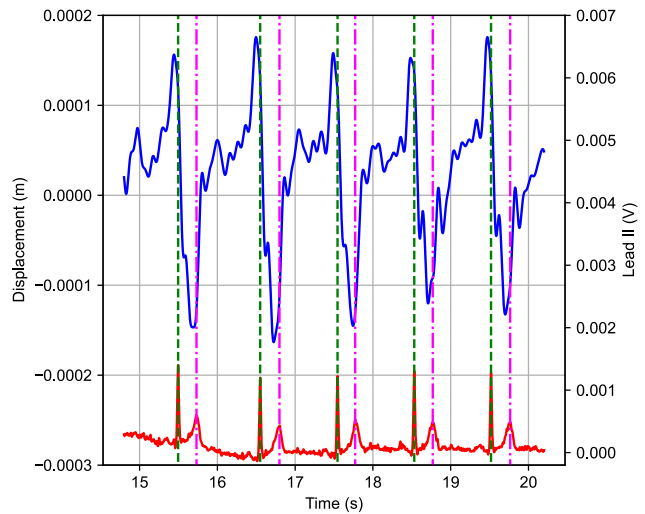

Fig. 22. Evolution of displacement with time, for the brightest range. The radar signal was synchronized with the ECG using the impulse experiment. The subject is the same as in Fig. 18. The superimposed vertical dashed green lines correspond to the times where the R-peaks are detected on the ECG (manual synchronization). The superimposed vertical dashed magenta lines correspond to the times where the T-peaks are detected on the ECG.

We can see the resulting velocity and displacement signals in Figs. 21 and 22. The observed results confirm our hypothesis: on the velocity signal, the R peak happens around the same time as the downward peak on the radar and the T peaks match the biggest upward peak of each pattern. Taking into account the whole captures made within this experiment, we measure an average delay of $48.0\,ms \pm 1.3\,ms$ between manual synchronization based on the correspondence of peaks and the signal synchronized using the impulse in the ECG. This difference might be explained by the delay between the depolarization or repolarization and the actual myocardium response [20]. This error is sufficiently low to base the synchronization algorithm described in the next subsection on the hypothesis described above. We also notice that, while always being present, the different peaks in the velocity signal may vary in amplitude (thus inducing a variation in the displacement signal). One hypothesis to explain these variations is that they might depend on some cardiac-related factors, for example arterial pressure. This hypothesis is not further discussed in this paper.

### C. Algorithm for Automated Synchronization

In this section we propose an algorithm to automate the synchronization of the ECG R-peaks with the tall downward peaks in the velocity signal. The first step is to detect the R-peaks on the ECG and the downward peaks on the radar velocity signal. For the former, we used the same technique based on clustering as described in section IV-B. For the latter, we used a *find_peaks* algorithm on the velocity signal. We then created a pseudo Dirac comb for each signal, each Dirac delta function corresponding to a detected peak. To create the pseudo-combs, we used the following formula, based on an approximation of a Dirac comb:

$$C(\{T_i\}_{i \in [\![1;n]\!]})(x) = \sum_{i=1}^{n} \frac{1}{|a|\sqrt{\pi}} \times \exp{-\left(\frac{x - T_i}{a}\right)^2} \quad (3)$$

where $\{T_i\}_{i \in [\![1;n]\!]}$ are the times corresponding to the heartbeats and $a$ is a coefficient that controls the width of the peaks in the curve. If $a \to 0$, each peak tends to a Dirac function. We can see in Fig. 23 an example of a pseudo-comb. We compute a point-to-point distance between the two combs and find the time shift (or offset) to apply to one of the signals which minimizes this distance. The offset corresponding to the smallest distance is the offset induced by the operator. There is a misalignment between the pseudo-combs of the ECG and the radar that is mostly caused by the difference in the frequency sampling of the ECG and radar signals. Thus, we need to choose a value of $a$ that guarantees that the peaks will still intersect even if they are not perfectly aligned. $a = 0.1$ meets these requirements. So, in our case, to find the offset induced by the operator, we apply an offset ranging from $-2$ s to $+2$ s with a step of 4 ms, to one of the signals and for each tested offset, we compute the distance between the pseudo-combs using Equation (4).

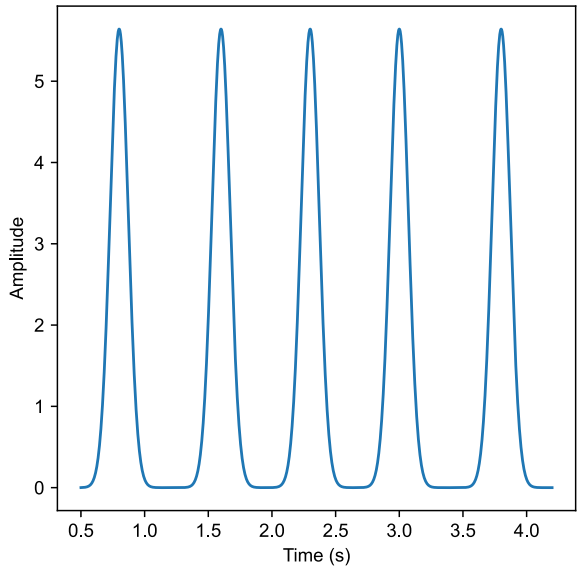

Fig. 23. Example of a pseudo-comb obtained using Equation 3. Each peak corresponds to the time of a heartbeat. $a = 0.1$.

$$d(C_{ECG}, C_{radar}) = \sum_i |C_{ECG}[i] - C_{radar}[i]| \quad (4)$$

where $C_{ECG}$ and $C_{radar}$ are the pseudo-combs generated from the ECG and the radar signals, respectively. The offset we are seeking is the one that produces the smallest distance. An example of the scores obtained by the algorithm for various shifts can be seen in Fig. 24. This algorithm is based on two assumptions:

- the operator does not induce a delay of more than 2 s between the captures of ECG and radar,

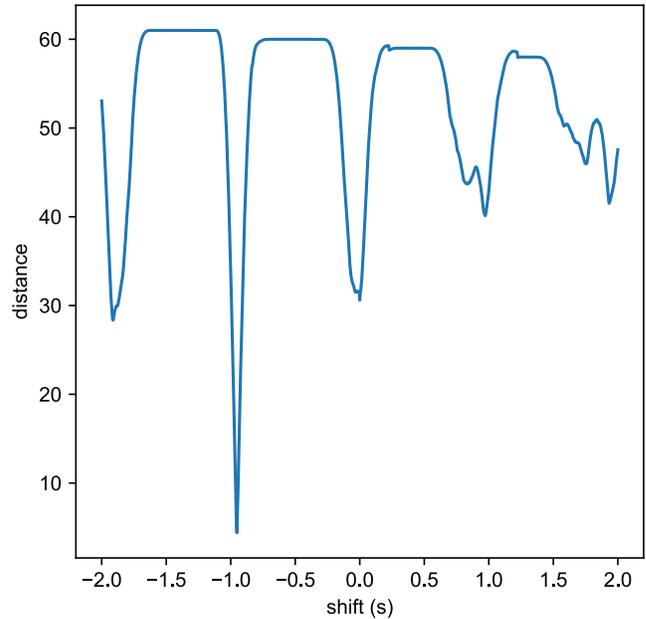

Fig. 24. Example of distance scores obtained for various shifts applied on the ECG signal. We can see that the shift -1 s corresponds to the lowest distance, thus the operator launched the ECG recording 1 s before the recording of the radar data.

- the heart rate is not totally regular: if the subject is breathing, there is some physiological variation of the heart rate induced by the breathing [24]. If the subject is not breathing, apnea makes the heart rate decrease over time [29].

If it is suspected that a bigger delay was induced by the operator, a bigger range of shifts can be tested by the algorithm. However, as more values are tested, the computing time will be longer.

## VII. CONCLUSION AND DISCUSSION

In this paper, we presented a data processing algorithm to extract, with a high accuracy, the micro chest displacements induced by the heart movements. We made a hypothesis based on the current knowledge in cardiology to synchronize the signals of the radar and the ECG and demonstrated it with an experiment. Hence we linked the cardiac movement to the observed skin movement it induced. Based on these results we proposed an algorithm to synchronize an ECG signal with the signal extracted from an FMCW radar. A concomitant analysis of FMCW radar sensor and echocardiography recordings would allow us to determine a more precise correlation between the heart mechanics and the observed skin movements. Such a study is currently ongoing. From a medical point of view, these results are interesting for several applications. The first one is that it is possible to accurately monitor the heart rate in a contactless way, which could help in the diagnosis of arrhythmia. But above all, heart monitoring with a FMCW radar could be useful for pathologies that are difficult to detect solely with the use of an ECG: the diseases that affect the contractility of the myocardium, like heart failure, are an example of such pathologies. By focusing on the mechanical activity of the heart rather than its electrical activity, FMCW radar combined with our data processing could provide a new tool to help establish a diagnosis, follow up a patient pathology and provide some prognostic elements. As we stated in the introduction, echocardiography delivers precise information about the heart kinetics but, as FMCW radar is far cheaper and easier to operate, it could be used as the first link in the chain of diagnosis. The experiments we described in this paper were performed on subjects with no known cardiac pathology. Therefore it would be interesting to reproduce them on people with known CVDs, so that some new criteria could be determined to interpret and capitalize on the signals described in this paper.